\documentclass[%reprint,
twocolumn,
nofootinbib,
 amsmath,amssymb,
 aps,
 longbibliography,
]{revtex4-2}

\usepackage[utf8]{inputenc} % allow utf-8 input
\usepackage[T1]{fontenc}    % use 8-bit T1 fonts
\usepackage{hyperref}       % hyperlinks
\usepackage{url}            % simple URL typesetting
\usepackage{booktabs}       % professional-quality tables
\usepackage{amsfonts}       % blackboard math symbols
\usepackage{nicefrac}       % compact symbols for 1/2, etc.
\usepackage{microtype}      % microtypography
\usepackage{graphicx}
\usepackage{xcolor, etoolbox}
\usepackage{amsmath,amsfonts,amsthm} % Math packages
\usepackage{mathtools}
\usepackage{physics}
\usepackage{ragged2e}
\usepackage{braket}      % Dirac Bra-Ket notation
\usepackage[english]{babel} % English language/hyphenation
\usepackage{enumitem}
\usepackage{xspace}
\usepackage{dsfont}
\usepackage{bm, bbm}% bold math
\usepackage{dcolumn}% Align table columns on decimal point
\usepackage[colorinlistoftodos,prependcaption,textsize=tiny]{todonotes}
\graphicspath{{./figures/}} % Set graphics path

\hypersetup{
    colorlinks,
    linkcolor=blue,
    citecolor=blue,
    urlcolor=blue,
    breaklinks=true 
}

\begin{document}

%----------------------------------------------------------------------------------------
%	TITLE SECTION
%----------------------------------------------------------------------------------------

\title{Relative entropic uncertainty relation}

\author{Stefan Floerchinger}
 \email{stefan.floerchinger@thphys.uni-heidelberg.de}
\author{Tobias Haas}
 \email{t.haas@thphys.uni-heidelberg.de}
\author{Ben Hoeber}
\email{hoeber@thphys.uni-heidelberg.de}
 \affiliation{Institut f\"{u}r Theoretische Physik, Universit\"{a}t Heidelberg, \\ Philosophenweg 16, 69120 Heidelberg, Germany}

%----------------------------------------------------------------------------------------
%	ABSTRACT SECTION
%----------------------------------------------------------------------------------------

\begin{abstract}
Quantum uncertainty relations are formulated in terms of relative entropy between distributions of measurement outcomes and suitable reference distributions with maximum entropy. This new type of entropic uncertainty relation can be applied directly to observables with either discrete or continuous spectra. We find that a sum of relative entropies is bounded from above in a non-trivial way, which we illustrate with some examples.
\end{abstract}

\maketitle

%----------------------------------------------------------------------------------------
%	DOCUMENT SECTION
%----------------------------------------------------------------------------------------

\section{Introduction}
The uncertainty relation lies at the heart of quantum theory and is also well known among laymen. It was introduced first by Heisenberg in 1927 \cite{Heisenberg1927} and proven in the same year by Kennard \cite{Kennard1927} (see also \cite{Weyl1928}). It states the simple but astonishing fact that position $x$ and momentum $k$ can not be fixed or measured with perfect precision at the same time. This idea was generalized to arbitrary hermitian operators $A$ and $B$,
\begin{equation}
    \sigma_A \, \sigma_B \ge \frac{1}{2} \abs{\braket{[A,B]}},
    \label{eq:RobertsonUR}
\end{equation}
where $\sigma_A$ denotes the standard deviation of $A$. Stated in this form, it becomes evident how the non-commutative nature of quantum mechanics is related to the uncertainty relation \cite{Robertson1929,Schroedinger1930}. Thus, another way of interpreting uncertainty relations, is that the order of subsequent measurements matters.

Although the uncertainty relation formulated in terms of standard deviations is often handy, it is nowadays believed that a formulation in terms of \textit{information entropies} is more adequate. One reason is that the standard deviation exhibits counterintuitive behavior in some situations \cite{Bialynicki-Birula2011}. Furthermore, entropy \textit{is} a measure for the total uncertainty of a probability distribution \cite{Shannon1948}. A detailed discussion about advantages of uncertainty relations using entropy instead of variances can be found in refs.\ \cite{Bialynicki-Birula2011,Bialynicki-Birula1984,Coles2017,Hertz2019,Floerchinger2020}.

The first entropic uncertainty relation (EUR) was formulated for the continuous variables position $x$ and conjugate momentum $k$. In terms of the corresponding probability densities $f(x)$ and $g(k)$ it reads \cite{Everett1957,Hirschman1957,Beckner1975,Bialynicki-Birula1975,Hertz2019}
\begin{equation}
    S(f) + S(g) \geq 1 + \ln \pi,
    \label{eq:BirulaEUR}
\end{equation}
where $S(f)$ denotes the differential entropy of the probability density $f(x)$. This entropic uncertainty relation is not only interesting because entropy is a more adequate measure of uncertainty, but also because for the position / momentum pair it implies and hence is stronger than Heisenberg's inequality. More precisely, it requires the distributions $f (x)$ and $g (k)$ to be of Gaussian form in order to saturate the bound, which can be seen by rewriting it in terms of variances again \cite{Son2015, Hertz2019}.

Later, an entropic uncertainty relation for (non-degenerate) observables $X$ and $Z$ with \textit{discrete} spectra
\begin{equation}
    \begin{split}
        X \ket{x} &= x \ket{x}, \\
        Z \ket{z} &= z \ket{z},
       \label{eq:EigenstatesDiscrete}
    \end{split}
\end{equation}
was found by Deutsch \cite{Deutsch1983}, improved by Maassen and Uffink \cite{Maassen1988} following a conjecture by Kraus \cite{Kraus1987}. Later, it was further strengthened by Berta \textit{et al.} \cite{Berta2010} (see also \cite{Frank2012}), resulting in
\begin{equation}
    S(p) + S(q) \geq \ln\frac{1}{c} + S(\rho).
    \label{eq:MaassenEUR}
\end{equation}
Therein $S(p)$ and $S(q)$ denote the classical Shannon entropies \cite{Shannon1948} of the discrete distributions $p(x)$ and $q(z)$, while the von Neumann entropy $S(\rho)$ accounts for the \textit{mixedness} of the system's state $\rho$. Moreover, $c$ is defined as the maximum overlap between any two eigenvectors of $X$ and $Z$
\begin{equation}
    c = \max_{x,z} \abs{\braket{x | z}}^2,
    \label{eq:MaximumOverlap}
\end{equation}
which quantifies what one may call the \textit{quantum incompatibility} of the two bases $\{ \ket{x}\}_x$ and $\{ \ket{z}\}_z$. 

The latter relation can also be generalized to two positive operator-valued probability measures (POVMs) $\{\Lambda_x\}_x$ and $\{ \Gamma_z \}_z$, which are non-negative operators that satisfy a completeness relation. Then, the Shannon entropies of the classical distributions $p (x) = \text{Tr} \{\rho \Lambda_x \}$ and $q (z) = \text{Tr} \{\rho \Gamma_z \}$ are bounded from below as prescribed by the Maassen-Uffink relation \eqref{eq:MaassenEUR}, if the quantum incompatibility gets replaced by \cite{Krishna2002,Tomamichel2012,Coles2014,Wilde2019}
\begin{equation}
    c = \max_{x,z} \norm{\sqrt{\Lambda_x} \, \Gamma_z \sqrt{\Lambda_x}}_{\infty} = \max_{x,z} \norm{\sqrt{\Lambda_x} \sqrt{\Gamma_z}}_{\infty}^2, 
    \label{eq:cPOVM}
\end{equation}
where $\norm{.}_{\infty}$ denotes the infinity operator norm. Note that for projective measurements, this operator norm reduces to the overlap again.

Until today many different types of EURs were investigated, for example in the context of Wehrl entropy \cite{Grabowski1984}, to classify the uncertainty between energy and time \cite{Hall2008, Boette2016, Hall2018, Coles2019} or for the family of Rényi entropies \cite{Maassen1988,Birula2006}. Moreover, the idea of uncertainty relations was also extended to bi- and tri-partite systems with quantum memory in terms of conditional entropies \cite{Renes2009, Berta2010, Tomamichel2011, Coles2012, Frank2013}. Also, strengthend versions of the bound were investigated, which further reduce the quantum incompatibility measure $c$ defined in \eqref{eq:cPOVM} \cite{Tomamichel2012,Coles2014}.

The two different types of observables were first unified in a formal way by Frank and Lieb \cite{Frank2012} (using the Golden-Thompson inequality and the Gibbs variational principle) based on previous work by Rumin \cite{Rumin2011}, who also generalized the relation to POVMs \cite{Rumin2012}. For the probability densities $f(x)$ and $g(k)$ their EUR reads
\begin{equation}
    S(f) + S(g) \ge \ln (2 \pi) + S(\rho),
    \label{eq:FrankLiebEUR}
\end{equation}
while for discrete variables it reduces to the result of Maassen and Uffink \eqref{eq:MaassenEUR}. Interestingly, it is tighter than \eqref{eq:BirulaEUR} for sufficiently mixed states and becomes tight in the infinite temperature limit $\beta \to 0$.

One of the main aims of this work is to present a simple EUR, which directly unifies the two cases of discrete and continuous random variables. In particular, we investigate a formulation in terms of \textit{relative entropy} instead of entropy, which allows to unify the two different types of observables in a straightforward manner.

Classically, the relative entropy (often also called Kullback-Leibler divergence) between two discrete distributions $p(x)$ and $\tilde{p}(x)$ is defined as \cite{Kullback1951,Kullback1959,Cover2006,Wilde2019}
\begin{equation}
    S(p \| \tilde{p}) = \sum_x p (x) \ln (p (x) / \tilde{p} (x)),
    \label{eq:RelativeEntropy}
\end{equation}
and can be seen as a measure of their \textit{distinguishability}. It can not be considered as an actual distance measure since it does not obey a triangle inequality and is not symmetric. In fact it is a \textit{divergence} in the mathematical sense.

Relative entropy plays a crucial role in the context of EURs. Some modern proofs of EURs for discrete variables (e.g. for the Maassen-Uffink relation \eqref{eq:MaassenEUR}) rely on the monotonicity property of the quantum version of the relative entropy under completely positive trace-preserving maps\footnote{This property is also often called \textit{data-processing inequality}.} \cite{Coles2011,Coles2012,Wilde2019}. Furthermore, EURs can also be stated as information exclusion principles in the presence of classical or quantum memory \cite{Hall1995,Hall1997,Coles2014}. In this sense, uncertainty between non-commuting observables is characterized by the impossibility of having a memory system that is strongly correlated to the measurements of both observables. Since mutual information can be expressed through relative entropy, such formulations can be regarded as EURs in terms of relative entropy in the presence of memory. 

Moreover, the concept of relative entropy was used in refs.\ \cite{Barchielli2017,Barchielli2018} to formulate \textit{measurement} EURs which deal with \textit{joint approximate} measurements of incompatible observables (cf. ref.\ \cite{Busch2014} for an overview, see also \cite{Coles2017}). In this field of research, EURs are used to characterize the quality of the aforementioned approximation or to constrain the disturbance of an observable after measuring another one (see e.g. \cite{Buscemi2014}), rather than to quantify the fundamental incompatibility of non-commuting observables, which is known as \textit{preparation} uncertainty instead. In this work we only deal with the latter type of uncertainty.

In contrast to these considerations, our aim is to reformulate existing EURs in terms of relative entropies. The main result of this work is the relative entropic uncertainty relation (REUR) \eqref{eq:REUR}, which provides a bound for a sum of two relative entropies from above. Besides adding a new perspective on EURs, the REUR is capable of describing the two cases of discrete and continuous variables simultaneously in one and the same inequality. This is  is achieved by picking suitable model distributions of maximum entropy $p_{\text{max}}$ and $q_{\text{max}}$. In the case of discrete variables, convenient choices would be the uniform or a generalized Boltzmann distribution. For continuous variables, the most suitable choice is the Gaussian distribution, which also allows for a strengthened reformulation of eq.\ \eqref{eq:RobertsonUR} highlighting the looseness of the latter for non-Gaussian distributions. In this sense, the REUR \eqref{eq:REUR} has to be understood as a reformulation of the Frank and Lieb EUR \cite{Frank2012} in terms of relative entropies allowing for a more direct unification of the two cases of variables. While the new formulation allows for a somewhat different physical interpretation, mathematically, it is equivalent to the relation proven in ref.\ \cite{Frank2012}.

The paper is organized as follows. We begin with introducing the relative entropy in more detail in \autoref{chap:Basics}. Then, we discuss the discrete case explicitly and show how to obtain an EUR in terms of relative entropy starting from Boltzmann distributions as optimal models in \autoref{chap:Discrete}. Afterwards, the continuous case is discussed in \autoref{chap:Continuous}, where convenient model distributions turn out to be Gaussians. In \autoref{chap:REUR}, we consider the general case and unify the relations obtained before. Furthermore, we make some remarks concerning the interpretation of the resulting EUR in terms of relative entropies. Also, we discuss a further non-trivial example, where we explicitly show that both sides of the REUR behave well when taking the continuum limit. Finally, we summarize our results and give an outlook in \autoref{chap:Conclusions}.

\paragraph*{Notation.}
 In this paper we adopt natural units, with $\hbar=1$. Operator hats are dropped completely. Instead, we use capital letters for operators $X$ and small letters for their eigenvalues $x$ and eigenvectors $\ket{x}$. As a consequence, random variables are denoted by small letters, too. Furthermore, we use the symbol $S$ for all kinds of entropies and take the underlying probability distribution $p(x)$, probability density $f(x)$ or density matrix $\rho$ as argument. In this sense it depends on the argument whether $S$ is a Shannon entropy $S(p) = - \sum_x p (x) \ln p (x)$, a differential entropy $S(f) = - \int d x f(x) \ln f(x)$ or a von Neumann entropy $S(\rho) = - \text{Tr} \{\rho \ln \rho \}$. It should be clear from the context what is meant in each case.

\section{Relative entropy}
\label{chap:Basics}
To get an intuitive understanding of relative entropy consider a random experiment where discrete events are correctly described by a distribution $p(x)$. If we mistakenly take the events to be distributed according to a model distribution $\tilde{p}(x)$, the relative entropy $S(p \| \tilde{p})$ as defined in eq. \eqref{eq:RelativeEntropy} quantifies the uncertainty deficit about $p(x)$ due to the wrong model $\tilde{p}(x)$, more precisely the difference between the cross entropy $S(p,\tilde{p}) = - \sum_x p (x) \ln \tilde{p} (x)$, which is the expectation value of the apparent information content, and the real information content $S(p) = - \sum_x p(x) \ln p (x)$ \cite{Vedral2002}.

Consequently, relative entropy is a non-negative quantity being zero if and only if the two distributions agree. Moreover, to have a finite value, it needs the support condition $\text{supp}(p(x))\subseteq \text{supp}(\tilde{p}(x))$ to hold, otherwise its value is set to $+ \infty$. The latter corresponds to the case where the model $\tilde{p}(x)$ predicts zero probability for events, which can in fact happen. In this case one can rule out the model $\tilde{p} (x)$ with certainty as soon as such an event occurs which causes the ``distinguishability''  $S(p \| \tilde{p})$ to be infinitely large. Thus, a suitable model $\tilde{p}(x)$ should at least cover all possible outcomes, otherwise it could be ruled out for particular events \cite{Cover2006}. Exchanging here $p (x)$ and $\tilde{p} (x)$ gives an intuitive understanding for the asymmetry of $S(p \| \tilde{p})$ under an exchange of its arguments.

It should be noted that the roles of the \textit{true} and the \textit{model} distribution can depend on the context. For our purposes it suffices to consider the second argument of relative entropy as the \textit{model}.

Interestingly, relative entropy has some crucial advantages over entropy. Most importantly it turns out to be well-defined for discrete and continuous random variables. Simply taking the continuum limit $p(x) \to f(x) dx$ and $\tilde{p} (x) \to \tilde{f} (x) d x$ yields \cite{Cover2006}
\begin{equation}
    S(f \| \tilde{f}) = \int d x \, f(x) \ln (f(x) / \tilde{f}(x)).
    \label{eq:RelativeEntropyContinuum}
\end{equation}
In contrast, Shannon's entropy does not yield the differential entropy when taking this continuum limit \cite{Jaynes1963}. As a consequence, differential entropy can become negative, while Shannon's entropy is a non-negative quantity. 

Moreover, relative entropy is invariant under a diffeomorphism $x \to x'(x)$ on the underlying statistical manifold. This is also not the case for the differential entropy, which can be seen easily by considering scaling transformations. Thus, it seems to be more appropriate to work with relative entropy instead of entropy, if one wants to describe discrete as well as continuous random variables.

\section{Discrete variables}
\label{chap:Discrete}
We now consider first the case of discrete variables explicitly. We will show that suitable models are Boltzmann distributions, which directly leads us to a reformulation of the Maassen-Uffink inequality \eqref{eq:MaassenEUR} in terms of relative entropies. Throughout the following section we assume the dimension of the Hilbert space to be finite $d = \dim \mathcal{H} < \infty$.

\subsection{Quantum-classical states and classical distributions}
We consider a quantum system in some arbitrary quantum state, which is specified by a non-negative, trace-class operator $\rho$ with normalization $\Tr \{\rho \} = 1$. We are interested in measuring two observables represented by hermitian trace-class operators $X$ and $Z$. Since we explicitly want to describe the discrete case, we assume that the spectra of both operators are bounded and discrete, such that the set of equations \eqref{eq:EigenstatesDiscrete} holds. Therefore, $\{\ket{x} \}_x$ and $\{\ket{z}\}_z$ are orthonormal bases with normalizable elements. 

The results of a measurement of $X$ are then distributed according to
\begin{equation}
    p(x) = \braket{x | \rho | x},
    \label{eq:ProbabilityProjectiveMeasurement}
\end{equation}
i.e. $p(x)$ is the probability to obtain the value $x$ when measuring $X$ in its eigenbasis. The described procedure is the \textit{projective measurement} (PVM), which is a special case of a POVM. In most generality, the probability distribution \eqref{eq:ProbabilityProjectiveMeasurement} can be constructed from the expectation value
\begin{equation}
    p(x) = \Tr \big\{ \Lambda_x \rho \big\},
\end{equation}
which reduces to Eq. \eqref{eq:ProbabilityProjectiveMeasurement} if the POVM elements are chosen to be $\Lambda_x = \ket{x} \bra{x}$, i.e. rank-1 projectors projecting the state $\rho$ onto the $x$-th eigenvalue of $X$.

In this setup, entropic uncertainty relations state that the sum of two classical entropies of the corresponding measured distributions $p(x)$ and $q(z)$ are bounded from below in a non-trivial way. 

Furthermore, one can define a quantum-classical state $\rho_X$ representing the state of the quantum system after measuring the operator $X$ without recording the result. More precisely, the projective measurement is mathematically represented by applying a measurement map $\mathcal{X}$ to the general density matrix $\rho$ as follows
\begin{equation}
    \rho \to \rho_X = \mathcal{X} (\rho) = \sum_{x} p (x) \ket{x} \bra{x}.
    \label{eq:ProjectiveMeasurement}
\end{equation}
Therein $p (x)$ is defined as in \eqref{eq:ProbabilityProjectiveMeasurement}, such that $\rho_X$ can indeed be interpreted as the classical state obtained after measuring the full quantum state $\rho$ in the eigenbasis $\{ \ket{x} \}_x$ without recording the result. Consequently, the von Neumann entropy of a measured density operator $\rho_X$ is the Shannon entropy of the classical probability distribution $p (x)$
\begin{equation}
    S(\rho_X)= - \sum_x p (x) \ln p (x) = S(p).
    \label{eq:ClassicalStateEntropy}
\end{equation}
Since projective measurements are unital quantum channels, we have the following relation to the full quantum entropy \cite{Nielsen2010}
\begin{equation}
    S (\rho) \leq S (p).
    \label{eq:UnitalCPTPMapEntropy}
\end{equation}
During a measurement without recording the result, the von Neumann entropy can increase but not decrease.

\subsection{Boltzmann distributions as optimal models}
To use relative entropy in a formulation of an EUR, we mainly need to specify the reference distributions $\tilde{p} (x)$ and $\tilde{q} (z)$. Depending on the availability of side information about the actual distributions $p (x)$ and $q(z)$, these distributions can be picked, such that they maximize an entropy. Let us make this idea more explicit by considering some examples in the following.

\subsubsection{No prior information}
In the absence of any constraints, the optimal model, i.e. the model with maximum Shannon entropy, is given by a uniform distribution
\begin{equation}
    \tilde{p} (x) = p_{\text{max}} (x) = 1/d.
\end{equation}
Then, the relative entropy between $p(x)$ and $p_{\text{max}} (x)$ reduces to a simple difference of entropies
\begin{equation}
    \begin{split}
        S (p \| p_{\text{max}}) &= - S(p) + \ln d \\
        &= - S(p) + S(p_{\text{max}}),
    \end{split}
\end{equation}
such that we can rewrite the Maassen-Uffink EUR \eqref{eq:MaassenEUR} as
\begin{equation}
    S (p \| p_{\text{max}}) + S (q \| q_{\text{max}}) \le - \ln \frac{1}{c} - S(\rho) + 2 \ln d.
    \label{eq:REURUniformModel}
\end{equation}
The sum of the divergencies of $p (x)$ from $p_{\text{max}} (x)$ and $q(x)$ from $q_{\text{max}} (x)$ is bounded from above. This bound becomes tighter the smaller the maximum overlap $c$ in eq. \eqref{eq:MaximumOverlap} is, and the larger the von Neumann entropy $S(\rho)$ is.

In the extreme case of a mutually unbiased basis (MUB) one has $\ln (1/c) = \ln d$ and for a maximally mixed state $S(\rho) = \ln d$. In this case the right hand side of eq. \eqref{eq:REURUniformModel} vanishes and neither $p(x)$ nor $q(x)$ can be distinguished from a uniform distribution with maximum entropy any more. This shows how the uncertainty relation has indeed an information theoretic significance in this formulation. 

\subsubsection{Given expectation values}
Another convenient choice of constraints is a set of fixed expectation values. This is in particular interesting if we consider thermal states or in general if we have some additional macroscopic information about a set of observables, which are distributed according to $p(x)$. For example, we can consider the expectation value of the true distribution $p(x)$
\begin{equation}
    \mu_x = \sum_x p (x) x
\end{equation}
and require that the reference distribution $\tilde{p} (x)$ has the same mean value $\tilde{\mu}_x = \mu_x$. Then, the optimal model in the sense that $S(\tilde{p})$ is maximal under the constraint $\tilde{\mu}_x = \sum_x \tilde{p} (x) x = \mu_x$ is given by
\begin{equation}
    \tilde{p} (x) = p_{\text{max}} (x) = \frac{1}{Z_x} e^{-\gamma_x x},
\end{equation}
where $\gamma_x$ is a Lagrange multiplier and $Z_x$ is a constant for proper normalization. If there exist a set of observables $Y_i$, which commute with $X$, then we can include their expectation values, too. For example, if one of these operators is the Hamiltonian $H$, one can consider a thermal state as the optimal model. In particular, the condition of equal energy expectation values for the distributions $p(x)$ and $\tilde{p} (x) = p_{\text{max}} (x)$ uniquely determines the inverse temperature $\beta$ of the thermal model. 

For a Boltzmann-type distribution $p_{\text{max}}(x)$ with equal expectation value we find for the relative entropy
\begin{equation}
    \begin{split}
        S (p \| p_{\text{max}}) &= - S(p) + \ln Z_x + \gamma_x \mu_x \\
        &= - S(p) + S(p_{\text{max}}),
    \end{split}
\end{equation}
i.e. it again reduces to a difference of entropies. If we consider two observables $X$ and $Z$ and choose Boltzmann-type distributions as optimal models, a reformulation of the Maassen-Uffink EUR \eqref{eq:MaassenEUR} reads
\begin{equation}
    \begin{split}
        &S(p \| p_\text{max}) + S(q \| q_{\text{max}}) \\
        &\le - \ln \frac{1}{c} - S(\rho) + \ln \left(Z_x \, Z_z \right) + \gamma_x \mu_x + \gamma_z \mu_z,
        \label{eq:REURDiscrete}
    \end{split}
\end{equation}
where we used indices for all quantities related to the two different observables $X$ and $Z$. These quantities are uniquely determined by the true distributions $p(x)$ and $q(z)$ and thus can be computed once the latter have been determined experimentally.

The latter considerations can be extended to the case where the mean values of the true distribution $p(x)$ and the model $\tilde{p} (x)$ are given, but do \textit{not} agree $\mu_x \neq \tilde{\mu}_x$. In this case, the relative entropy acquires an additional term, which is proportional to the difference of the two, 
\begin{equation}
    S(p \| p_{\text{max}}) = - S(p) + S(p_{\text{max}}) + \gamma_x (\mu_x - \tilde{\mu}_x).
\end{equation}
Nevertheless, the REUR \eqref{eq:REURDiscrete} remains the same.

\subsubsection{Given set of moments}
We can generalize the latter considerations to the case where an arbitrary set of moments is constrained. More precisely, we may consider $N$ functions $m_1(x), \ldots, m_N(x)$, for which the expectation values 
\begin{equation}
    \langle m_j \rangle = \sum_x p(x) m_j (x)
\end{equation}
are known. Thus, the optimal model distribution $\tilde{p} (x) = p_{\text{max}} (x)$ needs to fulfill the set of constraints
\begin{equation}
    \alpha_j (p, \tilde{p}) = \sum_x \left(\tilde{p}(x) - p (x) \right) m_j (x) = 0,
\end{equation}
for $j \in \{1,...,N \}$. Assuming the existence of such an optimal model $\tilde{p} (x)$\footnote{Note that in the case of continuous variables, solutions to this optimization problem do often not exist, especially when considering higher order moments.}, the maximum entropy principle dictates its form to be \cite{Jaynes1957,Jaynes1963}
\begin{equation}
    \tilde{p} (x) = p_{\text{max}} (x) = \exp \left(\sum_{j=0}^N \lambda_j \, m_j (x) \right),
    \label{eq:GeneralOptimalModelDiscrete}
\end{equation}
where $\lambda_j$ for $j>0$ are Lagrangian multipliers ensuring the $N$ constraints $\alpha_j (p, \tilde{p})=0$ and $\lambda_0$ is needed for proper normalization of the resulting optimal model distribution $\tilde{p} (x)$. Interestingly, the form of the solution \eqref{eq:GeneralOptimalModelDiscrete} does not change if the constraints are given in the form of inequalities $\alpha_j (p, \tilde{p}) \ge 0$, but one has now additional constraints $\lambda_j \ge 0$ for all $j \in \{1, ..., N \}$ for the optimization problem \cite{Rubinstein2004,Rubinstein2005}. 

\subsection{Formulation solely in terms of relative entropies}
Another interesting point is to rewrite \eqref{eq:REURDiscrete} entirely in terms of relative entropies. To that end we need to replace the quantum entropy $S(\rho)$ and the Shannon entropies of some optimal models by the quantum relative entropy, which is defined as \cite{Umegaki1962,Vedral2002}
\begin{equation}
    S(\rho \| \tilde{\rho}) = - \text{Tr} \{\rho \, (\ln \rho - \ln \tilde{\rho}) \}.
\end{equation}
It can be considered as the quantum analogue of \eqref{eq:RelativeEntropy} exhibiting the same properties on the level of density matrices $\rho$ and $\tilde{\rho}$.

As for the classical relative entropies, we consider models of maximum entropy. In particular, we construct the measured state $\rho_{X,\text{max}}$ corresponding to the optimal classical distribution
\begin{equation}
    \rho_{X, \text{max}} = \sum_x p_{\text{max}} (x) \ket{x} \bra{x},
\end{equation}
which can be done analogously for $\rho_{Z, \text{max}}$. Then, a quantum relative entropy of the actual state $\rho$ w.r.t. such a model can be written as
\begin{equation}
    S(\rho \| \rho_{X, \text{max}}) = - S(\rho) + S(p_{\text{max}}).
    \label{eq:RelativeEntropyQuantumClassical}
\end{equation}
Furthermore, we can use that the uniform distribution
\begin{equation}
    \rho_{\text{max}} = \frac{1}{d} \, \mathds{1},
\end{equation}
allows to rewrite the quantum entropy in a simple way
\begin{equation}
    S(\rho) = - S(\rho \| \rho_{\text{max}}) + \ln d.
\end{equation}
Then, a formulation of \eqref{eq:REURDiscrete} solely in terms of relative entropies reads
\begin{equation}
    \begin{split}
        &S(p \| p_\text{max}) + S(q \| q_{\text{max}}) \\
        &\le  \ln (c d) - S(\rho \| \rho_{\text{max}}) + S(\rho \| \rho_{X, \text{max}}) + S(\rho \| \rho_{Z, \text{max}}).
        \label{eq:REURDiscrete2}
    \end{split}    
\end{equation}
Note that for the maximum overlap as defined in eq.\ \eqref{eq:MaximumOverlap} we have $c\geq 1/d$ so that $\ln(c d) \geq 0$, with equality corresponding to mutually unbiased bases.

\section{Continuous variables}
\label{chap:Continuous}
Now we turn to continuous variables, and specifically consider position $X$ and momentum $K$ in one spatial dimension. We assume the corresponding probability density functions $f(x)$ and $g(k)$ to be supported everywhere on the real line, if not stated differently. 

Most importantly, the dimension of the Hilbert space is not finite any more and the optimal models of maximum entropy are described by probability density functions.

\subsection{Position, momentum and probability densities}
In contrast to discrete variables, we cannot assign a meaningful quantum-classical state $\rho_X$ after a measurement of $X$ to our quantum system of interest. Mathematically, this is due to the fact that the eigenstates of $X$ are not orthonormal (the same holds for $K$). Instead, their overlaps are given by Dirac deltas
\begin{equation}
    \braket{x | x'} = \delta (x-x') \text{   and   } \braket{k | k'} = \delta (k-k'),
\end{equation}
such that the formal extension of the projective measurement \eqref{eq:ProjectiveMeasurement} leads to an operator
\begin{equation}
    \rho \to \rho_X = \mathcal{X} (\rho) = \int d {x} \, f (x) \ket{x} \bra{x},
    \label{eq:measuredStateCont}
\end{equation}
which is not trace-class and thus cannot represent any quantum state. Nevertheless, one can still compute the density of the quantum state $\rho$ in the eigenbasis of the observable $X$, such that the probability distribution density $f(x) = \braket{x | \rho |x}$ (cf. eq. \eqref{eq:ProbabilityProjectiveMeasurement}) remains well-defined. Then, the probability for obtaining an outcome between $x$ and $x + d x$ is given by $f(x) d x$.

For the following discussion it is important to look at the scalar product of position and momentum eigenstates
\begin{equation}
    \braket{x | k} = \frac{1}{\sqrt{2 \pi}} e^{i x k},
\end{equation}
which reflects the fact that the spectra of the two operators $X$ and $K$ are related by a Fourier transform. Since the latter is true for all positions $x$ and momenta $k$, the maximum overlap $c$ is given by
\begin{equation}
    c = \max_{x,k} \abs{\braket{x | k}}^2 = \frac{1}{2 \pi}
\end{equation}
and the two bases $\{ \ket{x}\}_x$ and $\{ \ket{k}\}_k$ can be considered as mutually unbiased bases in a continuous sense.

Let us note here that every experimental measurement of a position $X$ or a momentum $K$ can only be carried out with finite accuracies $\delta x$ and $\delta k$ leading formally to discrete probability distributions and measurement outcomes in the form of histograms.

More precisely, actual measurement outcomes can be considered as being sampled from the underlying probability density $f(x)$, such that one ends up with a discrete probability $p_i$ to register for example a particle in the $i$th interval of size $\delta x$ given by \cite{Bialynicki-Birula2011,Coles2017}
\begin{equation}
    p_i = \int_{i \, \delta x}^{(i+1)\delta x} d x \, f(x).
\end{equation}
Consequently, one can associate a Shannon entropy to this distribution
\begin{equation}
    S (p) = - \sum_i p_i \ln p_i,
\end{equation}
which is a measure for the uncertainty about the measured distribution. In the limit of infinitely small bin sizes $\delta x \to 0$, the measured Shannon entropy $S(p)$ diverges to $+ \infty$ as a consequence of allowing for an infinite and thus unphysical precision. Therefore, the true differential entropy of the underlying probability density $S(f)$ can only be recovered in the limit $\delta x \to 0$ after subtracting this infinite additive constant \cite{Bialynicki-Birula2011,Coles2017}
\begin{equation}
    S(f) = \lim_{\delta x \to 0} \left(S(p) + \ln \delta x \right),
\end{equation}
which shows again that the differential entropy $S(f)$ is not the continuum limit of the Shannon entropy $S(p)$ and that it does not inherit all of the Shannon entropies' properties. Nevertheless, it is possible to estimate the differential entropy $S(f)$ from the measured probability distribution $p_i$ \cite{Beirlant1997}.

Some of these problems might be avoided or relaxed by working with relative entropies instead of entropies. Moreover, it is of course interesting by itself to formulate EURs in terms of differential relative entropies, as we will do in the following.

\subsection{Gaussian distributions as optimal models}
\label{sec:GaussianModels}
In principle, one can consider the same constraints as before to obtain optimal models. For given expectation values one would end up with distributions of Boltzmann type. Furthermore, if we put the quantum system into a box of finite length $L$, e.g. $x \in [-L/2, L/2]$, we can choose the uniform distribution as optimal model, for the case that we do not have any constraints to implement. In the general case of an unbounded interval $x \in (- \infty, \infty)$ this is not possible\footnote{Note that in the case of $x \in [-L/2, L/2]$ we still can have (discrete) $k \in (- \infty, \infty)$, such that the uniform model may not be chosen for both distributions at once.}.

Of special interest for the continuous case is a Gaussian distribution
\begin{equation}
    \tilde{f} (x) = f_{\text{max}} (x) = \frac{1}{\sqrt{2 \pi \sigma^2}} \exp \left(- \frac{(x- \mu)^2}{2 \sigma^2} \right),
    \label{eq:GaussianDistribution}
\end{equation}
which represents the optimal model, in the sense of a maximum differential entropy $S(f)$, for known variance
\begin{equation}
    \sigma^2_x = \int d x \, f (x) (x-\mu_x)^2
\end{equation}
and mean
\begin{equation}
    \mu_x = \int d x \, f (x) x.
\end{equation}
Computing the entropy of the Gaussian distribution \eqref{eq:GaussianDistribution} gives
\begin{equation}
    S(f_{\text{max}}) = \frac{1}{2} \ln \left(2 \pi e \sigma_x^2 \right),
    \label{eq:DiffEntropyGaussian}
\end{equation}
such that the reformulation of the Frank-Lieb EUR \eqref{eq:FrankLiebEUR} in terms of relative entropies reads
\begin{equation}
    \begin{split}
        &S(f \| f_\text{max}) + S(g \| g_{\text{max}}) \\
        &\le - S(\rho) + 1 + \ln \left(\sigma_x \sigma_k \right).
        \label{eq:REURContinuous}
    \end{split}
\end{equation}
Again the right hand side is fully determined by the true distributions $f(x)$ and $g(k)$.

Similarly, we can reformulate the EUR by Białynicki-Birula and Mycielski \eqref{eq:BirulaEUR},
\begin{equation}
    S(f \| f_\text{max}) + S(g \| g_{\text{max}}) \le \ln 2 + \ln \left(\sigma_x \sigma_k \right).
\end{equation}
This relation is particularly interesting if we express it in terms of variances again
\begin{equation}
    \sigma_x \, \sigma_k \ge \frac{1}{2} \, e^{S(f \| f_\text{max}) + S(g \| g_{\text{max}})}.
\end{equation}
Compared to Robertson's formulation \eqref{eq:RobertsonUR}, the latter uncertainty relation provides a stronger bound for non-Gaussian distributions, where non-Gaussianity is measured in terms of relative entropies. Hence, it becomes visible in which cases the EUR \eqref{eq:BirulaEUR} is stronger than the uncertainty relation \eqref{eq:RobertsonUR}. The latter relation was already discussed in \cite{Son2015} (see also \cite{Hertz2019}), where the relative entropy w.r.t optimal Gaussians $S(f \| f_{\text{max}})$ was interpreted as the neg-entropy $\mathcal{J} (f) = S(f_{\text{max}}) - S(f)$ accounting for non-Gaussianity\footnote{See also \cite{Genoni2008} for a discussion concerning non-Gaussianity of quantum states measured in terms of \textit{quantum} relative entropy.}. 

Finally, let us briefly comment on a possible rewriting of \eqref{eq:REURContinuous} entirely in terms of relative entropies as done in \eqref{eq:REURDiscrete2} for discrete variables. The strategy used in the discrete case breaks down for continuous variables due to the non-existence of a measured state in eq.\ \eqref{eq:measuredStateCont}. We did not find a way to circumvent this problem but will argue in section \ref{chap:REUR} that one arrives at a satisfactory formulation nevertheless.

\section{Uncertainty relation based on relative entropy}
\label{chap:REUR}
In the last section we generalize the two cases of discrete and continuous variables and provide a REUR which covers both cases. Therefore, in the following discussion we will write $p(x)$ and $q(z)$ for \textit{every} type of distribution and their precise meanings have to be deduced from the context. Furthermore, we discuss the resulting REUR in more detail.

\subsection{Model distributions of maximum entropy}
In the two previous sections we were mainly concerned with finding suitable model distributions $\tilde{p}(x)$ and $\tilde{q}(x)$, which turned out to be the key step for reformulating statements with entropy in terms of relative entropy (cf.\ also Refs.\ \cite{Haas2020a} and \cite{Haas2020b}).

For our purposes the most useful models $\tilde{p}(x)$ and $\tilde{q}(x)$ were the ones which allowed to write a relative entropy as a simple difference of entropies. Since in general relative entropy can be written as
\begin{equation}
    S(p \| \tilde{p}) = - S(p) + S(p, \tilde{p}),
    \label{eq:RelativeEntropyCrossEntropy}
\end{equation}
i.e. as the difference between a cross entropy\footnote{For continuous distributions the sum has to be replaced by an integral $\int dx$.} 
\begin{equation}
    S(p, \tilde{p}) = - \sum_x p (x) \ln \tilde{p} (x),
\end{equation}
and an entropy $S(p)$, we are interested in finding models $\tilde{p}(x)$ for which the cross entropy $S(p, \tilde{p})$ reduces to the entropy of the model $S(\tilde{p})$ under certain additional constraints, i.e.
\begin{equation}
    S(p, \tilde{p}) = S(\tilde{p})
    \label{eq:MaxEntropyCondition}
\end{equation}
for a set of constraints
\begin{equation}
    \alpha_j (p, \tilde{p}) = 0,
\end{equation}
where $j \in \{1, ..., N\}$ and $N$ is typically a small natural number. Examples for convenient constraints included equal standard deviation $\sigma_x = \tilde{\sigma}_x$ or equal expectation values of macroscopic quantities $\braket{O}_p = \braket{O}_{\tilde{p}}$. 

The condition \eqref{eq:MaxEntropyCondition} can be inserted into the relative entropy, which simply gives
\begin{equation}
    S(p \| \tilde{p}) = - S(p) + S(\tilde{p}).    
\end{equation}
Then, the non-negativity of the relative entropy is equivalent to
\begin{equation}
    S(p) \le S(\tilde{p}).    
\end{equation}
Since the latter inequality has to hold for all true distributions $p(x)$, the condition \eqref{eq:MaxEntropyCondition} requires that the model is \textit{optimal} $\tilde{p}(x) = p_{\text{max}}(x)$, in the sense that it maximizes the classical entropy under the given constraints.

Additionally, we can consider the case where the condition \eqref{eq:MaxEntropyCondition} is released below in the sense that we allow for terms linear in the constraints. This still leads to the distribution of maximum entropy, but it is less optimal in the sense that some of its properties do not coincide with those of the true distributions, for example they could have different mean values $\mu_x \neq \tilde{\mu}_x$.

Another advantage of optimal model distributions is that they often fulfill the support condition $\text{supp}(p(x))\subseteq \text{supp}(p_{\text{max}}(x))$ for all $x$, such that the relative entropy $S(p \| p_{\text{max}})$ will be finite. As a consequence, a sum of relative entropies is bounded from above in a trivial way
\begin{equation}
    S(p \| p_\text{max}) + S(q \| q_{\text{max}}) \le S(p_\text{max}) + S(q_\text{max}),
    \label{eq:TrivialBound}
\end{equation}
which can be seen from combining \eqref{eq:RelativeEntropyCrossEntropy} with \eqref{eq:MaxEntropyCondition} and using that
\begin{equation}
    S(p) + S(q) \ge 0,
\end{equation}
also holds for the continuous case. Formulating an entropic uncertainty relation in terms of relative entropies is then equivalent to finding a smaller bound than \eqref{eq:TrivialBound}.

\subsection{The uncertainty relation and discussion}
Denoting the model distributions with maximum entropy and equal constraints by $p_{\text{max}} (x)$ and $q_\text{max} (z)$ allows us to state the \textit{relative entropic uncertainty relation} (REUR) for the two true distributions $p(x)$ and $q(z)$ in its most general form
\begin{equation}
    \begin{split}
        &S(p \| p_\text{max}) + S(q \| q_{\text{max}}) \\
        &\le - \ln \frac{1}{c} - S(\rho) + S(p_\text{max}) + S(q_\text{max}),
        \label{eq:REUR}
    \end{split}
\end{equation}
where $c$ is again the maximum overlap between any two eigenstates in \eqref{eq:MaximumOverlap} for projective measurements, or the generalization in eq.\ \eqref{eq:cPOVM} for POVMs.

One has to be careful with the interpretation of this inequality because the usual logic is in some sense reversed. Therefore, a few comments are in order.

\paragraph{Sum of relative entropies is bounded from above.} 
The non-trivial bound on a sum of relative entropies comes from above and not from below. In addition, a sum of relative entropies is bounded from below by zero due to the non-negativity of all summands. Furthermore, it is bounded from above as shown in eq. \eqref{eq:TrivialBound}. The upper bound tells that the ability to distinguish the measurement outcome distributions $p(x)$ and $q(x)$ from models with minimal information is bounded. The discriminating power on the right hand side of \eqref{eq:REUR} gets smaller when the overlap between basis functions in eq. \eqref{eq:MaximumOverlap} gets smaller, or when the von Neumann entropy gets larger.

Being able to distinguish a distribution well from a model means to have a high amount of information additional to the prior information encoded in the model. Since all uncertainty relations are statements about never having all information about two non-commuting observables, this translates to an upper bound for a sum of relative entropies w.r.t. models representing maximal missing information.

\paragraph{Smaller bound from smaller overlap and mixedness.} 
The bound is smaller if the two observables have a smaller quantum incompatibility measured in terms of $c$. It is also smaller if the state of the system becomes more mixed due to the term $-S(\rho) \le 0$. Thus, \textit{quantum incompatibility} or \textit{mixedness} is encoded in having a \textit{small bound}. Note that the right hand side of \eqref{eq:REUR} becomes largest, and the bound correspondingly weakest, when $c \to 1$ and $S(\rho) \to 0$. This corresponds to the classical situation and absence of statistical fluctuations in the state $\rho$. Nevertheless, the discriminating power on the left hand side of \eqref{eq:REUR} is then still bounded, namely by the sum of maximum entropies (cf. eq. \eqref{eq:TrivialBound}).

\paragraph{Maassen-Uffink and Frank-Lieb relations follow as special cases.}
As we have seen in \autoref{chap:Discrete} and \autoref{chap:Continuous}, the REUR \eqref{eq:REUR} reduces to the Maassen-Uffink relation \eqref{eq:MaassenEUR} for discrete and to the Frank-Lieb relation \eqref{eq:FrankLiebEUR} for continuous variables, respectively. Therefore, the tightness of the REUR \eqref{eq:REUR} depends on the type of variables under consideration. For discrete variables, it is tight for mutually unbiased bases and states that are diagonal in either one of the two bases. In contrast, for continuous variables it becomes tight in the infinite temperature limit. Furthermore, as the REUR \eqref{eq:REUR} is mathematically equivalent to the measure-theoretic formulation by Frank and Lieb put forward in \cite{Frank2012}, it could be proven in complete analogy using the Golden-Thompson inequality and Gibbs variational principle. 

\paragraph{Discrete and continuous spectra are unified.}
The main advantage of the REUR \eqref{eq:REUR} is that the left hand side is well defined in both cases, i.e. for operators with discrete as well as continuous spectrum, which is one of the nice features of relative entropy. On the right hand side we still have a sum of entropies, but for optimal models $p_{\text{max}} (x)$ and $q_{\text{max}} (z)$ we can always express these in terms of quantities appearing in the constraints $\alpha_j$. Thus, the relation is indeed free of ambiguities.

\paragraph{Bound contains additional knowledge.}
In most cases, the bound of an uncertainty relation depends on some measure of quantum incompatibility (either a commutator or an overlap) and the state $\rho$. Interestingly, the bound in the REUR \eqref{eq:REUR} does also depend on the additional knowledge encoded in the optimal models $p_{\text{max}} (x)$ and $q_{\text{max}} (z)$, i.e. on the constraints. Therefore, any available information about the distributions can be implemented directly in the presented uncertainty relation. This feature may be of particular interest for experimental applications where different kinds of constraints may be accessed.

\paragraph{Change of normalization.}
Specifically for continuous variables it is instructive to study scaling transformations of the form $| x \rangle \to | x^\prime \rangle = \alpha | x \rangle$. For $|\alpha|^2 \neq 1$ they change the normalization of the basis $\{ \ket{x}\}_x$. 
One needs to complement this change of basis with a change of integration measure $dx \to dx^\prime = |\alpha|^{-2} dx$ such that the probability
\begin{equation}
\langle x | \rho | x \rangle dx \to \langle x^\prime | \rho | x^\prime \rangle dx^\prime = \langle x | \rho | x \rangle dx
\end{equation}  
remains unchanged. The differential entropy changes then according to $S(p_\text{max})\to S(p_\text{max}) - \ln(|\alpha|^2)$. For the maximum overlap as defined in \eqref{eq:MaximumOverlap} one has $c\to c^\prime = |\alpha|^2 c$ such that $-\ln(1/c) \to - \ln(1/c^\prime) = - \ln(1/c) +\ln(|\alpha|^2)$. 
Taken together, the right hand side of \eqref{eq:REUR} is invariant under such a ``change of normalization'' transformation. When $|z\rangle$ corresponds to a continuous variable there is a similar invariance.

\subsection{Example of angle and angular momentum}
We close our analysis by considering a non-trivial example, namely angular momentum states and corresponding discrete or continuous angles. We will demonstrate here how the bound of the REUR \eqref{eq:REUR} behaves if we start with operators with discrete spectra and take the continuum and infinite volume limits. Note that the limits have to be understood in a formal way and that the left hand side of \eqref{eq:REUR} is always well defined as explained around \eqref{eq:RelativeEntropyContinuum}.

We start from a finite set of $2J+1$ spin or angular momentum eigenstates $|m \rangle$ with $L_z |m \rangle = m |m \rangle$ and where $m\in \{ -J, \ldots, J\}$ is an integer. As usual, $J$ can be integer or half-integer. Following ref.\ \cite{Barnett1990} we introduce now the angle states
\begin{equation}
| \phi \rangle = \frac{1}{\sqrt{2J+1}} \sum_{m=-J}^J e^{-im\phi} | m \rangle.
\end{equation}
In particular one has with this definition $|\phi \rangle = | \phi + 2\pi \rangle$ and 
\begin{equation}
e^{-i\varphi L_z} | \phi \rangle = | \phi + \varphi \rangle,
\end{equation}
as it should be. Because there is a continuum of angle states, they are overcomplete and have the overlap
\begin{equation}
\langle \varphi | \phi \rangle = \frac{\sin\left((J+\tfrac{1}{2})(\varphi-\phi)\right)}{(2J+1) \sin\left(\tfrac{1}{2}(\varphi-\phi)\right)}.
\end{equation}
In this regard the angle states are similar to coherent states. Note also the completeness relation
\begin{equation}
\frac{2J+1}{2\pi}\int_{0}^{2\pi} d\phi \, | \phi \rangle \langle \phi | =  \sum_{m=-J}^J |m\rangle \langle m | = \mathbbm{1}.
\label{eq:completenessAngles}
\end{equation}
One can therefore understand a measurement of the continuous angle $\phi$ as a POVM.

It is also possible to restrict to a discrete set of angle states $|\theta_j\rangle$ with
\begin{equation}
\theta_j = \theta_0 + \frac{2\pi j}{2J+1}, \quad\quad j=0, 1, \ldots, 2J. 
\label{eq:anglesTheta}
\end{equation}
These discrete angle states are actually orthonormal, $\langle \theta_j | \theta_k \rangle=\delta_{jk}$. In the continuum limit $J\to \infty$ the corresponding angles become dense but the states need then to be normalized as continuum states. It is also possible to introduce position variables on the circle $x=2\pi R \phi$, conjugate momenta $k=L_z / (2\pi R)$, and to consider an ``infinite volume'' limit $R\to \infty$.

Now let us discuss our relative entropic uncertainty relations in this context. First, for a measurement of the angular momentum $L_z$ with outcome $m$ one can use in the case of finite $J$ any of the reference distributions discussed in section \ref{chap:Discrete}. For example, if no information is available the uniform distribution would be a sensible reference, and if $\langle m \rangle$ and $\langle m^2 \rangle$ are known, a sensible reference distribution would be of the form
\begin{equation}
    q_\text{max}(m) = \exp(\lambda_0+\lambda_1 m + \lambda_2 m^2),
\label{eq:pmaxmGaussian}
\end{equation}
with Lagrangian multipliers $\lambda_j$. The corresponding entropy appearing on the right hand side of eq.\ \eqref{eq:REUR} is simply $S(q_\text{max}) = - \lambda_0 - \lambda_1 \langle m \rangle - \lambda_2 \langle m^2 \rangle$. The Gaussian reference distribution \eqref{eq:pmaxmGaussian} remains (for $\lambda_2 < 0$) also normalizable in the continuum limit $J\to \infty$ and it corresponds to a continuous Gaussian distribution of momenta $k=m/(2\pi R)$ in the infinite volume limit $R\to \infty$ when the corresponding variances are held fixed. 

For a measurement of the angle the situation is more involved. First, for finite $J$ one can consider the projective measurement in the basis $|\theta_j\rangle$ corresponding to the angles \eqref{eq:anglesTheta}. This is then a discrete observable which can be treated similar to $L_z$ above. Note that the overlap as defined in eq.\ \eqref{eq:MaximumOverlap} is minimal, $c=\max_{m,j}|\langle m | \theta_j \rangle|^2=1/(2J+1)$. 

A measurement of the continuous angle states $|\phi \rangle$ is instead a POVM. Because the range of possible values is finite, the uniform distribution $p_\text{max}(\phi)=1/(2\pi)$ with $S(p_\text{max})=\ln(2\pi)$ is an admissible reference distribution with maximum entropy when no additional information is given. Another interesting reference distribution is the von Mises distribution \cite{Mardia2009}
\begin{equation}
    p_\text{max}(\phi) = \frac{e^{\kappa \cos(\phi-\mu)}}{2\pi I_0(\kappa)},
\end{equation}
where $I_i (\kappa)$ is the modified Bessel function of the first kind and of order $i$. It corresponds to a maximum entropy distribution for $\phi$ under the condition that the first circular moment
\begin{equation}
    \langle e^{i\phi} \rangle = \frac{I_1(\kappa)}{I_0(\kappa)} e^{i\mu},
\end{equation}
is known. The magnitude and complex phase of the latter fix the parameters $\kappa\geq 0$ and $0\leq \mu < 2\pi$. The differential entropy is in this case given by 
\begin{equation}
    S(p_\text{max}) = \ln(2\pi I_0(\kappa)) - \kappa \frac{I_1(\kappa)}{I_0(\kappa)},
    \label{eq:diffEntropyMises}
\end{equation}
and the quantum incompatibility $c$ as defined in eq.\ \eqref{eq:cPOVM} evaluates to $c=1/(2\pi)$. We have dropped here a multiplicative term $d\phi$ which would appear from a direct application of eq.\ \eqref{eq:completenessAngles} in eq.\ \eqref{eq:cPOVM} but cancels with a similar term that arises in the transition from a Shannon entropy to the differential entropy in eq.\ \eqref{eq:diffEntropyMises}. This fixes all state independent terms on the right hand side of the REUR \eqref{eq:REUR}.

Let us note that for large $\kappa$, the von Mises distribution becomes strongly localized around $\phi=\mu$ and approaches a Gaussian shape. In this sense one can recover the Gaussian models discussed in section \ref{sec:GaussianModels} in the infinite volume limit $R\to \infty$ for fixed variance of position $x$. 

We also note that up to the different normalization of states, the two possibilities to look at angle measurements (discrete as PVM and continuous as POVM) become equivalent in the continuum limit $J\to \infty$.

\section{Conclusion and Outlook}
\label{chap:Conclusions}
In summary, we have investigated a formulation of an entropic uncertainty relation in terms of \textit{relative entropies}. In particular, we found that a sum of relative entropies of the two true distributions with respect to model distributions of maximum entropy are bounded from above in a non-trivial way. More precisely, there is a bound depending on the entropies of the optimal models, which gets reduced by \textit{quantum incompatibility} in terms of the maximum overlap $c$ and \textit{mixedness} in terms of the von Neumann entropy $S(\rho)$ of the quantum state $\rho$. The main advantage of the presented formulation was that it allowed to cover the cases of observables with discrete and continuous spectra within one and the same entropic uncertainty relation. This was due to the fact that relative entropy, in contrast to Shannon entropy, behaves well when considering the continuum limit.

We have illustrated the benefits of this formulation on the specific example of finite and infinite spaces of angular momentum states and the conjugate angle variables. All state-independent quantities appearing in the uncertainty relation can then be easily evaluated and behave also favorable under the continuum as well as infinite volume limits.

For the future it might be interesting to generalize the presented approach to situations with quantum memory.

Another interesting point for future work concerns the application of the (relative) entropic uncertainty relation to quantum systems where entropies exhibit (unphysical) infinities while relative entropies remain finite. What we have in mind here are specifically applications to quantum field theory in terms of (functional) relative entropies. Such a formulation would allow to quantify entropic uncertainty e.\ g.\ for a scalar quantum field and its conjugate momentum field.

\section*{Acknowledgements}
This work is supported by the Deutsche Forschungsgemeinschaft (DFG, German Research Foundation) under Germany's Excellence Strategy EXC 2181/1 - 390900948 (the Heidelberg STRUCTURES Excellence Cluster), SFB 1225 (ISOQUANT) as well as FL 736/3-1.

%----------------------------------------------------------------------------------------
%	REFERENCES SECTION
%----------------------------------------------------------------------------------------

\bibliography{references.bib}

\end{document}